\documentclass[12pt]{article}

\usepackage{amssymb}
\usepackage{amsmath}
\usepackage{mathrsfs}

\begin{document}
\thispagestyle{empty}

\begin{flushright}
 KEK-TH-1127, 
 OIQP-06-20
\end{flushright}

\vspace{10mm}
\begin{center}
{\Large \bf Quantum Anomalies at Horizon \\
 and \\
 Hawking Radiations in Myers-Perry Black Holes}

\vspace{10mm}
{\large
Satoshi Iso$^a$\footnote{E-mail address: satoshi.iso@kek.jp},
Takeshi Morita$^a$\footnote{E-mail address: tmorita@post.kek.jp}
and
Hiroshi Umetsu$^b$\footnote{E-mail address: hiroshi\_umetsu@pref.okayama.jp}
}

\vspace{10mm}
$^a$ {\it Institute of Particle and Nuclear Studies, High Energy Accelerator
Research Organization(KEK),  
Oho 1-1, Tsukuba, Ibaraki 305-0801, Japan} \\
$^b$ {\it Okayama Institute for Quantum Physics,
Kyoyama 1-9-1, Okayama 700-0015, Japan}

\end{center}

\vspace{10mm}
\begin{abstract}
A new method has been developed recently to derive Hawking radiations from
black holes based on considerations of gravitational and gauge anomalies at
the horizon \cite{RW}\cite{IUW1}.  In this paper, we apply the method to
Myers-Perry black holes with multiple angular momenta in various dimensions
by using the dimensional reduction technique adopted in the case of
four-dimensional rotating black holes \cite{IUW2}.
\end{abstract}

\newpage
\setcounter{page}{1}
\setcounter{footnote}{0}

\baselineskip 6mm

\section{Introduction}
\setcounter{equation}{0}

Hawking radiation is the quantum effect to arise for
quantum fields in a background space-time with an event horizon.
There are many different derivations, from the
original calculation based on Bogoliubov transformations \cite{Hawking}
to Euclidean approaches \cite{Euclidean}, and all of them universally 
give the same answer. 
The universality tells us that the Hawking radiation 
must be determined  only by some universal 
quantum effects just on the horizon.\footnote{
The effect of scatterings away from the
horizon changes the spectrum to gray.  But this is not
the universal part of the Hawking radiation and we do not 
discuss it in the present paper.}

In the seminal paper by Robinson and Wilczek \cite{RW}, it was proposed that
the flux of Hawking radiation can be fixed by the amount of the
gravitational anomaly at the horizon. The method was then generalized in
\cite{IUW1} to charged black holes by using the gauge anomaly in addition to
the gravitational anomaly and further applied to rotating black holes
\cite{IUW2}\cite{soda} and others \cite{others}.  The essential observation
in \cite{RW} is that quantum fields near the horizon behave as an infinite
set of two-dimensional fields and ingoing modes at the horizon can be
considered as left moving modes while outgoing modes as right moving modes.
Once the ingoing modes fall into the black hole, they never come out
classically and cannot affect the physics outside the black hole.  Quantum
mechanically, however, they cannot be neglected because, without the ingoing
modes, the theory becomes chiral at the horizon, which makes the effective
theory anomalous under general coordinate or gauge transformations.  In this
sense, the ingoing modes at the horizon only affect the exterior region
through quantum anomalies. This is the basic idea of the method but there is
a slight difference between the original calculation in \cite{RW} and that
in \cite{IUW1}.  In this paper, we adopt the calculation used in
\cite{IUW1}.  We will explain the difference in the appendix.

This approach is similar to the beautiful derivation of the Hawking flux
based on conformal anomalies \cite{CF}, in which the Hawking flux was
obtained by solving the conservation law of the energy-momentum tensor with
the information of conformal anomaly.  In our method, instead of conformal
anomaly, we use gravitational or gauge anomalies.  Both anomalies are
quantum effects but there are the following differences. The gravitational
or gauge anomalies arise only for chiral theories while conformal anomaly
can arise even for vector-like theories. The reason that the gravitational
and gauge anomalies are relevant to the Hawking radiation is due to the
chiral decomposition property near the horizon.  Namely, quantum fields near
the horizon can be decomposed into the left (ingoing) and right (outgoing)
modes and the left modes are causally decoupled from the exterior physics
classically.  Also the gravitational and gauge anomalies are independent of
the details of quantum fields and can be universally determined.  Another
advantage to use the gauge anomalies is that we can derive the fluxes of
charges and angular momenta in addition to the energy flux.

In this paper, we apply the method to higher-dimensional Myers-Perry (MP)
black holes \cite{MP}.  Hawking flux from MP black holes with a single
charge (angular momentum) was obtained in \cite{soda} but they could not
obtain the flux in cases with multiple angular momenta.  In this paper, by
using similar dimensional reduction technique adopted in \cite{IUW2}, we
show that the anomaly method can be similarly applied to MP black holes with
more than one charge, and that it reproduces the flux of each angular
momentum $F_{ai} $ and that of energy-momentum tensor $F_{M} $ associated
with the Hawking radiation by each partial wave,
\begin{eqnarray}
 F_{ai} &=& \int_0^\infty \frac{d\omega}{2\pi}  
  m_i\left( N_{\{m_j\}}(\omega)- N_{\{-m_j\}}(\omega)\right) 
  = \frac{m_i}{2\pi}
  \sum_{j=1}^n \frac{m_j a_j}{r_H^2 + a_j^2}, \nonumber \\
 F_{M} 
  &=&  \int_0^\infty \frac{d\omega}{2\pi}  
  \omega \left(N_{\{m_j\}}(\omega) + N_{\{-m_j\}}(\omega)\right)
  = \frac{1}{4\pi}\left(\sum_{i=1}^n \frac{m_i a_i}{r_H^2 + a_i^2}\right)^2 
  + \frac{\pi}{12\beta^2}.
\end{eqnarray}
Here $N_{\{m_i\}}(\omega)$ is the Planck distribution with 
an inverse-temperature $\beta$ and chemical potentials 
$a_i/(r_H^2+a_i^2)$ 
for the angular momenta $m_i ~(i=1,2,\cdots, n)$.\footnote{
Here we used the Planck distribution for fermions in order to avoid the
problem of superradiance which is related to scatterings away from the
horizon.} 
These fluxes are given by the sums of contributions from a particle with
charge $\{m_i\}$ and antiparticle with charge $\{-m_i\}$.
%

The organization of the paper is as follows.  In section 2, we consider
quantum fields in the background of the MP black holes in various dimensions
and show that they behave as an infinite set of two-dimensional conformal
fields near the event horizon.  In section 3, we consider symmetries for
them and relate the conservation laws of the original energy-momentum tensor
to those in the dimensionally reduced theories.  In section 4, we obtain the
Hawking fluxes based on two-dimensional gravitational and gauge anomalies.
Section 4 is devoted to discussions.  In appendix, we explain the difference
between the original calculation in \cite{RW} and that in \cite{IUW1}.

\section{Quantum fields in Myers-Perry black hole}
\setcounter{equation}{0}

In the $D=2n+1+\epsilon ~(\epsilon = 0 \mbox{ or } 1)$ dimensions, 
the metric of the Myers-Perry black hole is given by \cite{MP}
\begin{eqnarray}
 ds^2 &=& dt^2 - \epsilon r^2 d\alpha^2 
  - \sum_{i=1}^n (r^2+a_i^2)
  \left(d\mu_i^2+\mu_i^2d\phi_i^2\right)
  - \frac{\mu r^{2-\epsilon}}{\Pi F}
  \left(dt - \sum_{i=1}^n a_i \mu_i^2d\phi_i\right)^2 \nonumber \\
  && - \frac{\Pi F}{\Pi - \mu r^{2-\epsilon}}dr^2, 
 \end{eqnarray}
where 
\begin{eqnarray}
 F &=& 1-\sum_{i=1}^n \frac{a_i^2\mu_i^2}{r^2+a_i^2}, \\
 \Pi &=& \prod_{i=1}^n (r^2 + a_i^2).
\end{eqnarray}
The following constraint is imposed for $\mu_i ~(i=1, 2, \cdots,n)$ and
$\alpha$,
\begin{eqnarray}
 \sum_{i=1}^n \mu_i^2 + \epsilon\alpha^2 = 1, 
  \qquad (0\leq \mu_i \leq 1, ~-1 \leq \alpha \leq 1).
\end{eqnarray}
This metric describes a black hole space-time with the mass
$M=\frac{(D-2)A_{D-2}}{16\pi G}\mu$ and angular momenta $\frac{2}{D-2} M
a_i$ in the $\phi_i$-directions, where $A_{D-2}$ is the volume of $S^{D-2}$.
This black hole is stationary and has $U(1)^n$ isometries with the Killing
vectors $\partial_{\phi_i}$.  We assume the existence of horizons located at
positive solutions of $\Pi - \mu r^{2-\epsilon}=0$. The inverse of
metric is given by
\begin{eqnarray}
 && g^{tt} = \frac{(\Pi-\mu r^{2-\epsilon})F 
  + \mu r^{2-\epsilon}}{(\Pi-\mu r^{2-\epsilon})F}, \nonumber \\
 && g^{t\phi_i} = \frac{\mu r^{2-\epsilon}}{(\Pi-\mu r^{2-\epsilon})F}\cdot
  \frac{a_i}{r^2+a_i^2}, \nonumber \\
 && g^{\phi_i\phi_j} = -\frac{1}{\mu_i^2 (r^2+a_i^2)} \delta_{ij}
  + \frac{\mu r^{2-\epsilon}}{(\Pi-\mu r^{2-\epsilon})F}\cdot
  \frac{a_ia_j}{(r^2+ a_i^2)(r^2+a_j^2)}, \nonumber \\
 &&  g^{rr} = - \frac{\Pi -\mu r^{2-\epsilon}}{\Pi F}, \nonumber \\
 && g^{\mu_i\mu_j} = - \frac{1}{r^2+a_i^2}\delta_{ij} 
       + \frac{r^2}{F}\cdot
       \frac{\mu_i \mu_j}{(r^2+ a_i^2)(r^2+a_j^2)},
\end{eqnarray}
and the determinant by
\begin{eqnarray}
 \label{detg}
 \sqrt{|g|} = \frac{\Pi F}{r^{1-\epsilon}} \sqrt{\gamma_{D-2}},
\end{eqnarray}
where $\sqrt{\gamma_{D-2}}$ is the determinant of the metric of $S^{D-2}$,
\begin{eqnarray}
 \sqrt{\gamma_{D-2}}=
  \left\{
   \begin{array}{lc}
    {\displaystyle \prod_{i=1}^{n-1} \mu_i} &  \mbox{for } D=2n+1, \\
    {\displaystyle \frac{1}{|\alpha|}\prod_{i=1}^{n} \mu_i} 
     &  \mbox{for } D=2n+2.
   \end{array}
  \right.
\end{eqnarray}

We consider a scalar field $\varphi$ in the Myers-Perry black hole
background. The action is
\begin{eqnarray}
 S = \frac{1}{2}\int d^Dx \sqrt{|g|}g^{\mu\nu}\partial_\mu \varphi
  \partial_\nu \varphi + S_{int}, 
\end{eqnarray} 
where $S_{int}$ includes a mass term and interaction terms.
Near the outer horizon, which is located at $r=r_H$ the largest root of 
$\Pi - \mu r^{2-\epsilon}=0$, the kinetic term gives a dominant
contribution to the action and thus we can ignore a mass and interaction
terms $S_{int}$. Hence the action becomes near the outer horizon as
\begin{equation}
 S = - \frac{1}{2}\int dtdr d\Omega_{D-2}(\mu r)
  ~\varphi
  \left[
   \frac{\Pi}{\Pi-\mu r^{2-\epsilon}}
   \left(\partial_t + \sum_{i=1}^n
    \frac{a_i}{r^2+a_i^2}\partial_{\phi_i}\right)^2
    -\partial_r \frac{\Pi-\mu r^{2-\epsilon}}{\Pi}\partial_r
  \right]
  \varphi.
\end{equation}
Note that the terms including $\partial_{\mu_i} \varphi$ in the kinetic part
are also suppressed near the horizon compared to the above terms.  
The scalar field $\varphi$ can be expanded by the spherical harmonics 
$Y^{R}_{m_1 \cdots m_n}(\mu_i, \phi_i) $ on $S^{D-2}$, where $R$ is a
label of representations of $SO(D-1)$ and  
$i\partial_{\phi_i} Y^{R}_{m_1 \cdots m_n}(\mu_i, \phi_i) 
= m_i Y^{R}_{m_1 \cdots m_n}(\mu_i, \phi_i)$.
Performing the expansion, 
$\varphi = \sum_{R, m_i} \varphi^R_{m_1 \cdots m_n}(t,r) 
Y^{R}_{m_1 \cdots m_n}(\mu_i, \phi_i)$,
the physics near the horizon can be effectively described by an infinite
collection of massless $(1+1)$-dimensional fields with the following action,  
\begin{equation}
 S = - \int dtdr (\mu r)
  ~(\varphi^R_{m_1 \cdots m_n})^*
  \left[
   \frac{\Pi}{\Pi-\mu r^{2-\epsilon}}
   \left(\partial_t + \sum_{i=1}^n
    \frac{im_i a_i}{r^2+a_i^2}\right)^2
   - \partial_r \frac{\Pi-\mu r^{2-\epsilon}}{\Pi}\partial_r
  \right]
  \varphi^R_{m_1 \cdots m_n}.
\end{equation}

From this action we find that $\varphi^R_{m_1 \cdots m_n}$ can be considered
as a $(1+1)$-dimensional complex scalar field in the background of the
dilaton $\Phi$, metric $g_{\mu\nu}$ and $U(1)$ gauge fields
$A^{(\phi_i)}_\mu$,
\begin{eqnarray}
 \label{2d-bg}
 && \Phi = \mu r, \qquad 
  g_{tt} = \frac{\Pi-\mu r^{2-\epsilon}}{\Pi} \equiv f(r), \qquad
  g_{rr} = - \frac{1}{f(r)}, \qquad 
  g_{tr} = 0, \nonumber \\
 && A^{(\phi_i)}_t = - \frac{a_i}{r^2+a_i^2}, \qquad
  A^{(\phi_i)}_r = 0.
\end{eqnarray}
The partial wave $\varphi^R_{m_1 \cdots m_n}$ has the $U(1)$ charge $m_i$
for each gauge field $A^{(\phi_i)}_\mu$.

\section{Symmetries and conservation laws}
\setcounter{equation}{0} 
When we derive Hawking radiations from anomalies at
the horizon in the next section, we need to use various Ward-Takahashi
identities in the two-dimensional effective theories.  In this section, we
derive them by reinterpreting the original conservation laws associated with
general coordinate transformations in $D=2n+1+\epsilon$ dimensions.

Following the general procedure in the Kaluza-Klein compactification, we
write the $D$-dimensional metric $g_{AB} ~(A, B = t, r, \mu_i, \phi_i)$ for
the Meyers-Perry black holes in terms of a $d$-dimensional metric
($d=n+1+\epsilon$), $U(1)^n$ gauge fields and dilatons,
\begin{align}
(g_{AB}) =
\begin{pmatrix}
g_{\alpha\beta} + h_{\phi_i \phi_j} A^{(\phi_i)}_\alpha A^{(\phi_j)}_\beta &
h_{\phi_j \phi_k} A^{(\phi_k)}_\alpha \\
h_{\phi_i \phi_k} A^{(\phi_k)}_\beta & h_{\phi_i \phi_j} \\
\end{pmatrix},
\end{align} 
where the indices $\alpha$ and $\beta$ denote $d$-dimensional coordinates
$(t, r, \mu_i) $ and $\phi_i$'s are the angular coordinates.  Note that
since the metric $g_{AB}$ does not depend on these angular coordinates
there are $U(1)_{\phi_i}$ isometries.  Then the $D$-dimensional general
coordinate transformation generated by a Killing vector
$\xi^{\phi_i}(t,r,\mu_i)$ becomes $U(1)_{(\phi_i)}$ gauge transformation in
the $d$ dimensions.  By using the Ward-Takahashi identity with respect to
this transformation, we can obtain a conservation law for $U(1)$ gauge
current.  Under this transformation, the fields change as
\begin{align}
\delta A_\alpha^{ (\phi_i)} = \partial_\mu \xi^{\phi_i},\quad 
\delta g_{\alpha\beta} =\delta h_{\phi_i \phi_j} =  0,~
\end{align} 
and the partition function in the above background changes as
\begin{align}
 \int d^{D}x  \delta g_{AB} \frac{\delta}{\delta g_{AB}} 
 \ln Z[g] 
 &= i\int d^Dx \xi^{\phi_i} \partial_\alpha 
 \left( \sqrt{|g|} T^\alpha_{\phi_i} \right) \nonumber \\
 &=i\int d^{D}x \xi^{\phi_i} 
 \left[
 \partial_r \left(\sqrt{|g|} T^r_{\phi_i} \right) 
 + \sum_{i=j}^{n-1+\epsilon}
 \partial_{\mu_j} \left( \sqrt{|g|} T^{\mu_j}_{\phi_i}\right) 
 \right] = 0.
\label{EM-phi}
\end{align} 
Here 
$\displaystyle{
T_{AB} = -\frac{i}{\sqrt{|g|}}\frac{\delta}{\delta g^{AB}} \ln Z[g]
}$ 
is a
$D$-dimensional energy-momentum tensor and depends only on $r$ and $\mu_i$
due to the isometries of the background. We define an $r$-component of a
$U(1)$ gauge current in the two-dimensional field theory as
\begin{equation}
 \label{Jr}
 J_{(\phi_i)}^r (r) \equiv 
  -\int d\Omega_{D-2} \left(\frac{\Pi F}{r^{1-\epsilon}}\right)
  T^r_{\phi_i}.
\end{equation}
We assume that on the right hand side in eq. (\ref{EM-phi}), the second term
is negligibly small compared to the first term near the horizon of the MP
black hole. This corresponds to the fact that the terms including
$\partial_{\mu_i}\varphi$ could be dropped in the discussion of the
effective action near the horizon in the previous section.  We then can see
from eq.(\ref{EM-phi}) that the following conservation law holds near the
horizon,
\begin{equation}
 \partial_r J_{(\phi_i)}^r = 0.
\end{equation}
Namely, by studying the behavior of the two-dimensional $U(1)$ gauge
current, we can see the flux of the $D$-dimensional angular momentum.  

Next, by considering a $D$-dimensional general coordinate 
transformation generated by a Killing vector 
$\xi^{t}(t,r,\mu_i)$,   we can obtain another conservation law.
Variations of the fields under it are
\begin{align}
 & \delta g_{\alpha\beta} =  g_{\beta t}  \nabla_\alpha   \xi^t+
 g_{\alpha t}  \nabla_\beta   \xi^t, \quad
 \delta A_\alpha^{ (\phi_j)} = \xi^t \nabla_t  
 A_\alpha^{ (\phi_j)} + A_t^{ (\phi_j)} \nabla_\alpha \xi^t, 
 \nonumber \\
 & \delta h_{\phi_i \phi_j} =  \xi^t \partial_t  h_{\phi_i \phi_j} =0,
\end{align} 
where $\nabla_\alpha$ is the covariant derivative associated with the
$d$-dimensional metric $g_{\alpha\beta}$.
Then a change of the partition function is given by
\begin{align}
 &\int d^{D}x   \delta g_{AB} \frac{\delta}{\delta g_{AB}} \ln Z[g]  
 \nonumber \\
 & = i\int d^{D}x \xi^{t} 
 \left[\partial_\alpha \left( \sqrt{|g|}
 \left(T^\alpha_{t} - A^{(\phi_i)}_t T^\alpha_{(\phi_i)}\right)\right)
 - \sqrt{|g|} \sum_{i=1}^n F^{(\phi_i)}_{t\alpha} T^{\alpha}_{\phi_i}
 \right] \nonumber \\
 &=i\int d^{D}x \xi^{t} \left[\partial_r 
 \left(\sqrt{|g|} \left(T^r_t - A^{(\phi_i)}_t T^r_{\phi_i}\right)\right) 
 +\partial_{\mu_i} \left(\sqrt{|g|} 
 \left(T^{\mu_i}_t - A^{(\phi_j)}_t T^{\mu_i}_{\phi_j}\right)\right) 
 - \sqrt{|g|}\sum_{i=1}^nF^{(\phi_i)}_{t\alpha} T^{\alpha}_{\phi_i}
 \right] \nonumber \\
 & = 0,
 \label{EM-t}
\end{align}
where we used the isometries of the MP backgrounds. 
$F^{(\phi_i)}_{t\alpha}$ are defined by the
components of the field
strengths of the $U(1)$ gauge fields $A^{(\phi_i)}_\alpha$,
\begin{eqnarray}
 && F^{(\phi_i)}_{t\alpha} 
  = \partial_t A^{(\phi_i)}_\alpha - \partial_\alpha A^{(\phi_i)}_t, \\
 && A^{(\phi_i)}_t 
  = - \frac{\mu r^{2-\epsilon}}{(\Pi - \mu r^{2-\epsilon})F+\mu r^{2-\epsilon}}
  \cdot \frac{a_i}{r^2 + a_i^2}, \\
 && A^{(\phi_i)}_{\mu_j} = A^{(\phi_i)}_r = 0. 
\end{eqnarray}
The equation (\ref{EM-t}) holds in the whole region outside the horizon
but  it can be simplified near the horizon as follows.
Near the horizon, because of $\Pi - \mu r^{2-\epsilon}=0$, 
$F^{(\phi_i)}_{t\mu_j}$
vanishes and $F^{(\phi_i)}_{tr} = -\partial_r A^{(\phi_i)}_t$ 
where the potential becomes
\begin{equation}
 A^{(\phi_i)}_t \longrightarrow - \frac{a_i}{r^2 + a_i^2}, 
  \qquad (r\longrightarrow r_H).
\end{equation}
These gauge field backgrounds, of course, coincide with those in
eq. (\ref{2d-bg}) derived by dimensional reduction of the action near the
horizon. Furthermore we assume that in eq. (\ref{EM-t}),
$\partial_{\mu_i}\left(\sqrt{|g|} 
\left(T^{\mu_i}_t - A^{(\phi_j)}_t T^{\mu_i}_{\phi_j}\right)\right)$ 
is negligible compared
to the other terms near the horizon, by a similar reason to the one
described below eq. (\ref{Jr}).  Defining a component $T^r_{t(2)}(r)$ of the
two-dimensional energy-momentum tensor as
\begin{equation}
 T^r_{t(2)}(r) \equiv \int d\Omega_{D-2} 
  \left(\frac{\Pi F}{r^{1-\epsilon}}\right) 
  \left(T^r_t - A^{(\phi_i)}_t T^r_{\phi_i}\right),
\end{equation}
we find  from eq. (\ref{EM-t}) the following conservation equation near
the horizon, 
\begin{equation}
 \partial_r T^r_{t(2)} - \sum_{i=1}^n F^{(\phi_i)}_{rt} J_{(\phi_i)}^r =0.
  \label{WT-T}
\end{equation}
The energy flux in $D$ dimensions is given by the two-dimensional energy
current. The second term in the above Ward-Takahashi identity can be
interpreted as the dissipation of energy due to interactions of the current
$J^r_{(\phi_i)}$ with the background electric field $F_{rt}^{(\phi_i)}$.

\section{Quantum anomalies and Hawking fluxes}
\setcounter{equation}{0}

Ingoing modes near the horizon are classically
irrelevant to physics outside the horizon. If we neglect the ingoing
modes, in the two-dimensional effective theory near the horizon gauge
and diffeomorphism invariance are broken by quantum anomalies. The
underlying theory is, of course, invariant. Therefore these anomalies
are cancelled by quantum effects of ingoing modes which are irrelevant
classically. In the following we show that the condition for vanishing
of anomalies at the horizon leads to the correct Hawking fluxes of
angular momenta and energy.

First we consider the $U(1)$ currents $J^r_{(\phi_i)}$ in the effective
theory which is defined in $r \in [r_H, \infty]$.  Each current corresponds
to angular momentum in the $\phi^i$-direction in the $D$-dimensional theory,
as shown in the previous section. We divide the region outside the horizon
into two regions, $r \in [r_H, r_H + \epsilon]$ near the horizon and $r >
r_H + \epsilon$.  If we omit the ingoing modes near the horizon, $U(1)$
current has gauge anomaly there. The consistent form of the abelian gauge
anomaly~\cite{Bardeen,Bertlmann,Fujikawa} in two dimensions is given by
\begin{equation}
 \nabla_\mu J^\mu_{(\phi_i)} 
  = - \frac{m_i}{4\pi\sqrt{-g_{(2)}}}
  \epsilon^{\mu\nu} \partial_\mu {\cal A}_\nu,
  \label{gaugeanomaly1}
\end{equation}
where $\epsilon^{01}=+1$ and $\mu, \nu$ run $t$ and $r$.
In our case, ${\cal A}_\mu$ is a sum of $n$ $U(1)$ gauge fields multiplied
by charges, 
 \begin{eqnarray}
  && {\cal A}_\mu = \sum_{j=1}^n m_j A^{(\phi_j)}_\mu, \\
  && {\cal A}_t = - \sum_{j=1}^n \frac{m_j a_j}{r^2 + a_j^2}, 
   \qquad {\cal A}_r = 0.
\end{eqnarray} 
It is noted
that the charge of the partial mode in the effective theory is
$m_i$. The consistent anomaly satisfies the Wess-Zumino condition. The
consistent current is derived from the quantum effective action and is
not gauge covariant. We can define covariant current as
\begin{equation}
 \tilde{J}_{(\phi_i)}^\mu \equiv J_{(\phi_i)}^\mu 
  - \frac{m_i}{4\pi\sqrt{-g_{(2)}}} \epsilon^{\mu\nu} {\cal A}_\nu, 
\end{equation}
which satisfies
\begin{equation}
\nabla_\mu \tilde{J}_{(\phi_i)}^\mu 
 = -\frac{m_i}{4\pi\sqrt{-g_{(2)}}}\epsilon^{\mu\nu} {\cal F}_{\mu\nu}.
 \label{gaugeanomaly2}
\end{equation}
Here ${\cal F}_{\mu\nu}$ is given by
\begin{eqnarray}
 {\cal F}_{\mu\nu} = \partial_\mu {\cal A}_\nu - \partial_\nu {\cal A}_\mu
  = \sum_{i=j}^n m_j F^{(\phi_j)}_{\mu\nu}.
\end{eqnarray}
In the case we consider here, the consistent current
$J^\mu_{(\phi_i),H}$ in $r \in [r_H, r_H+\epsilon]$ depends only on $r$
and satisfies the following anomalous equation with the gauge
field background (\ref{2d-bg}),
\begin{equation}
 \partial_r J^r_{(\phi_i),H} =  
  \frac{m_i}{4\pi} \partial_r {\cal A}_t.
\end{equation}
In the outside region $r > r_H + \epsilon$, the current is conserved, 
$\partial_r J^r_{(\phi_i),O} = 0 $.
Hence we can solve them in each region as
\begin{eqnarray}
 && J^r_{(\phi_i),O} = c^i_O, \\
 && J^r_{(\phi_i),H} = c^i_H 
  + \frac{m_i}{4\pi}\left({\cal A}_t(r) - {\cal A}_t(r_H)\right),
\end{eqnarray}
where $c^i_O$ and $c^i_H$ are integration constants. $c_O^i$ is the value
of the current at $r \rightarrow \infty$ and $c_H^i$ is the value of the
consistent current of the outgoing modes at the horizon $r=r_H$.  The
current is written as a  sum in two regions
\begin{equation}
 J^{\mu}_{(\phi_i)} = J^\mu_{(\phi_i),O}\Theta_+(r) + J^\mu_{(\phi_i),H}H(r),
\end{equation}
where $\Theta_+(r) = \Theta(r - r_H - \epsilon)$ and
$H(r)=1-\Theta_+(r)$ are step functions which are defined in the region
$r \geq r_H$.  Note that this current is a part of the total current
because we omitted the ingoing modes near the horizon. The total current
including a contribution of the ingoing modes is given by
\begin{equation}
 J_{(\phi_i)total}^\mu = J_{(\phi_i)}^\mu + K_{(\phi_i)}^\mu,
\end{equation}
where
\begin{equation}
\label{K}
 K_{(\phi_i)}^r = - \frac{m_i}{4\pi}{\cal A}_t(r) H(r).
\end{equation}
This contribution cancels the anomalous part of $J_{(\phi_i)}^\mu$ near
the horizon.

In the previous papers \cite{RW}\cite{IUW1}\cite{IUW2}, the invariance of
the quantum effective action under gauge and general coordinate
transformations was considered. 
Requiring the invariance can determine the
relation between the integration constants $c_O^i$ and $c_H^i.$ This method
has revealed the significance of each contribution from the ingoing and
outgoing modes, but the relation itself can be easily obtained by imposing
the conservation of the total current, $\partial_r J_{(\phi_i)total}^r = 0$.
This condition gives the relation,
\begin{equation}
 c^i_O = c^i_H - \frac{m_i}{4\pi}{\cal A}_t (r_H).
\end{equation}


In order to fix the value of the current, we need a further
boundary condition at the horizon.
Here we impose that the covariant form of the outgoing
current should vanish at the horizon. 
The outgoing modes  can produce possible divergences
for physical quantities seen by infalling observers into black holes, 
as discussed in, for example, \cite{Unruh} \cite{CF}.
The physical vacuum in the black hole backgrounds
should be chosen as such that these unphysical divergences
should vanish. 
In our case, this condition corresponds to the above boundary
condition.
We discuss it in the appendix again.
Another condition to specify the vacuum is the 
boundary condition for the ingoing currents $K_{(\phi_i)}^\mu$.
In eq. (\ref{K}) we have already chosen an integration constant
such that it vanishes at $r \rightarrow \infty$. 
This condition corresponds to taking Unruh vacuum, 
in stead of Hartle-Hawking vacuum.

Since the covariant current is given by $\tilde{J}_{(\phi_i)}^r =
J_{(\phi_i)}^r + \frac{m_i}{4\pi}{\cal A}_t(r)H(r)$, the condition
$\tilde{J}_{(\phi_i)}^r(r_H) = 0$ determines the value of the charge
flux at $r \rightarrow \infty$ as
\begin{equation}
 c^i_O = - \frac{m_i}{2\pi}{\cal A}_t(r_H)
  = \frac{m_i}{2\pi} \sum_{j=1}^n \frac{m_j a_j}{r_H^2 + a_j^2}.
\end{equation}
This coincides with the flux of angular momentum associated with the
Hawking radiation.

Next we consider the flux of energy-momentum radiated from the
Myers-Perry black holes.  Omitting the ingoing modes in the near horizon
region $r_H \leq r \leq r_H + \epsilon$, the consistent energy-momentum
tensor $T^\mu_{\nu,H}$ satisfies a modified conservation equation with
the consistent anomalies.  In two dimensions the energy-momentum tensor
for right-hand modes satisfies the following  Ward-Takahashi identity with
$U(1)$ gauge fields $A^{(\phi_i)}_\mu$ and dilaton $\Phi$ backgrounds,
\begin{equation}
 \nabla_\mu T^\mu_\nu = 
  \sum_{i=1}^n 
  \left(F^{(\phi_i)}_{\mu\nu}J^\mu_{(\phi_i)} 
   + A^{(\phi_i)}_\nu \nabla_\mu J_{(\phi_i)}^\mu
  \right)
  - \frac{\partial_\nu \Phi}{\sqrt{-g_{(2)}}}
  \frac{\delta S}{\delta \Phi}
  + {\mathscr A}_\nu,
\end{equation}
where ${\mathscr A}_\mu$ is the consistent gravitational
anomaly~\cite{Alvarez} which is given by
\begin{equation}
 {\mathscr A}_\mu = \frac{1}{96\pi\sqrt{-g_{(2)}}}
  \epsilon^{\nu\rho}\partial_\rho \partial_\sigma
  \Gamma^\sigma_{\mu\nu}.
\end{equation}
This energy-momentum tensor is not covariant under general coordinate
transformations.  On the other hand the covariant energy-momentum
$\tilde{T}^\mu_\nu$ satisfies the Ward-Takahashi
 identity with the same form as the
above but the anomaly term ${\mathscr A}_\mu$ is replaced with the covariant
one $\tilde{{\mathscr A}}_\mu = - \frac{1}{96\pi\sqrt{-g_{(2)}}}
\epsilon_{\mu\nu}\nabla^\nu R$.

In the case considered here, the $\nu=t$ component of the Ward-Takahashi
identity in the consistent form becomes
\begin{equation}
\label{EM-consistent}
 \partial_r T^r_{t,H} = \sum_{i=1}^n
  \left(F^{(\phi_i)}_{rt}J^r_{(\phi_i)} 
   + A^{(\phi_i)}_t \nabla_\mu J_{(\phi_i)}^\mu
  \right)
  + \partial_r N^r_t(r),	
\end{equation}
where background fields (\ref{2d-bg}) are used. $N^r_t(r)$ is defined
by ${\mathscr A}_t = \partial_r N^r_t$,
\begin{equation}
 N^r_t(r) = \frac{1}{192\pi}\left(f'^2(r) + f(r) f''(r)\right).
 \label{gravanomaly1}
\end{equation}
Eq. (\ref{EM-consistent}) is the same Ward-Takahashi identity as 
eq. (\ref{WT-T}) if there are no anomalous terms. 
See \cite{IUW1} for the Ward-Takahashi identity in presence of anomalies. 
In eq. (\ref{EM-consistent}), the first and second terms in the
right hand side are combined in terms of the covariant current
$\tilde{J}_{(\phi_i)}^r$ as $F^{(\phi_i)}_{rt} \tilde{J}_{(\phi_i)}^r$.
By substituting 
$\tilde{J}_{(\phi_i)}^r = c^i_O + \frac{m_i}{2\pi}{\cal A}_t(r)$
into the equation, $T^r_{t,H}$ is obtained as
\begin{eqnarray}
 T^r_{t,H} = a_H + \int^r_{r_H}dr \partial_r
  \left(
   -\frac{1}{2\pi}{\cal A}_t(r_H){\cal A}_t(r)
   +\frac{1}{4\pi}{{\cal A}_t}^2(r)
   + N^r_t(r)
  \right).
\end{eqnarray}
On the other hand, the energy-momentum tensor $T^r_{t,O}$ in the outside
region $r > r_H$ satisfies 
\begin{equation}
 \partial_r T^r_{t,O} = \sum_{i=1}^n F^{(\phi_i)}_{rt} J_{(\phi_i),O}^r.
\end{equation}
By using $J_{(\phi_i),O}^r = c^i_O$ this is solved as
\begin{equation}
 T^r_{t,O} = a_O -\frac{1}{2\pi} {\cal A}_t(r_H) {\cal A}_t(r).
\end{equation}
The energy-momentum tensor combines contributions from these two
regions, $T^\mu_\nu = T^\mu_{\nu,O}\Theta_+ + T^\mu_{\nu, H}H$.
This does not contain a contribution from the ingoing modes near the
horizon. The total energy-momentum tensor is a sum of $T^\mu_\nu$ and
$U^\mu_\nu$, where
\begin{equation}
 \label{EM-in}
 U^r_t = -\left(
	   \frac{1}{4\pi} {\cal A}_t^2(r)
	   + N^r_t(r)
	  \right)H(r)
\end{equation} 
is a contribution from the ingoing modes. To determine a constant part
of $U^r_t$, we require the condition that the current should vanish at
$r \rightarrow \infty$. This condition corresponds to 
vanishing ingoing energy
flow at $r \rightarrow \infty$.

We can again obtain a relation between 
the integration constants  $a_O$  and $a_H$
from the conservation of total energy-momentum tensor
\begin{equation}
 a_O = a_H 
  + \frac{1}{4\pi}{{\cal A}_t}^2(r_H) - N^r_t(r_H).
\end{equation}


In order to determine the flux of energy $a_O$ at $r\rightarrow \infty$,  
we impose a vanishing condition 
for the covariant energy-momentum tensor 
at the horizon, $\tilde{T}^r_t (r_H)=0$. This corresponds to the
regularity condition for the energy-momentum tensor at the future
horizon. In this case, the covariant anomaly is given by 
$\tilde{{\mathscr A}}_t = \partial_r \tilde{N}^r_t$ where
\begin{equation}
 \tilde{N^r_t} =
  \frac{1}{96\pi}\left(ff''-\frac{1}{2}(f')^2\right),
  \label{gravanomaly2}
\end{equation}
and the covariant energy-momentum tensor is related to the consistent one
as 
\begin{equation}
 \tilde{T}^r_t = T^r_t + \frac{1}{192\pi}
  \left( ff'' - 2 (f')^2 \right).
\end{equation}
Therefore $a_H$ is determined as 
\begin{equation}
a_H = \frac{\kappa^2}{24\pi} = 2N^r_t(r_H),
\end{equation}
where the surface gravity $\kappa$ at the horizon is
\begin{equation}
 \kappa = \frac{2\pi}{\beta} = \frac{1}{2}f'(r_H)
  = \frac{\Pi'(r) - (2-\epsilon)\mu r^{1-\epsilon}}{2\mu r^{2-\epsilon}}
  \Bigg|_{r=r_H}.
\end{equation}
The flux of energy-momentum tensor is given by
\begin{eqnarray}
 a_O &=& \frac{1}{4\pi}{{\cal A}_t}^2(r_H) 
  + N^r_t(r_H) \nonumber \\
 &=& \frac{1}{4\pi} 
  \left(
   \sum_{i=1}^n \frac{m_i a_i}{r_H^2 + a_i^2}
  \right)^2 
  + \frac{\pi}{12\beta^2}.
\end{eqnarray}
Note that the flux is independent of the 
dimension $D$ of the space-time. 
In the non-rotating cases ($a_i=0$), the Hawking
flux is proportional to $(T_H)^2$ where
$T_H=1/\beta$ is the Hawking temperature of the
black hole. In $D$ dimensions the 
energy density of the black body radiation with temperature
$T_H$ is proportional to $(T_H)^D$ (Stephan-Boltzmann law.)
Since the area of black holes is proportional to
${\cal A} \sim M^{D-2} \sim (T_H)^{2-D}$,
the total flux behaves as $(T_H)^2$, which is like the
two-dimensional  Stephan-Boltzmann law.

\section{Discussions}
\setcounter{equation}{0}

In this paper we have applied the method of quantum anomalies to derive the
Hawking flux from Myers-Perry black holes with multiple angular momenta in
various dimensions.  The method adopted here made only use of the quantum
anomalies at the horizon and in this sense it is very universal. Namely, it
does not depend on the details of the quantum fields away from the horizon.
But we have obtained only the total flux of energy or charges and a natural
question is whether we can obtain more detailed information about the black
body radiation from black holes.

Black body spectrum of the Hawking radiations with the Hawking temperature
is deformed by the gray body factor due to the effect of scatterings away
from the horizon.  This is, of course, not universal and we need more
information about the black hole background away from the horizon.  But the
radiation before it is modified may be universally given and there is a
chance that we can obtain the black body spectrum based on the quantum
anomalies.  Near the horizon, quantum fields behave as an infinite set of
two-dimensional conformal fields.  Such fields have infinitely many
conserved currents with higher spins in addition to the energy momentum
tensor or gauge currents.  In curved space-times, they will acquire quantum
anomalies if the fields are chiral, and values of the anomalies for these
higher spin currents can determine the black body spectrum of the Hawking
radiations.  We would like to report it in our future publication
\cite{IMU1}.

Another issue is the entropy of black holes.  Although the black hole
entropy is well understood macroscopically, its microscopic understanding is
yet incomplete.  Since black hole entropy is also universally determined as
the Hawking radiation, we may expect that it can be given only by some
universal quantum effects at the horizon.  The black hole entropy may be
given by the number of degrees of freedom of the ingoing modes of some
gravitational modes at the horizon (which are classically irrelevant but
quantum mechanically important to physics outside the black holes).  We
calculated the entropy based on the idea and the result is proportional to
the area, but the coefficient is slightly different from the
Bekenstein-Hawking entropy.  Since the entropy must be universal as well as
the Hawking radiation, we believe that our approach for the quantum physics
near the horizon will be important to investigate the thermodynamic
properties of black holes.  We would like to further study and report it in
near future.

\vspace{5mm}
After completing this work we found a preprint \cite{Xu} in which Hawking
radiations in general Kerr-(anti)de Sitter black holes are studied.

\section*{Acknowledgements}
T. M. was supported in part by a JSPS Research Fellowship for Young
Scientists.

\section*{Appendix: What is different between \cite{RW} and \cite{IUW1,IUW2}}
\setcounter{equation}{0}

Since there seem to be some confusions among the readers of \cite{RW} and
\cite{IUW1, IUW2}, we will explain the difference between them here.
Although the basic ideas are essentially the same, there are the following
differences in the calculations.

 In the paper \cite{RW}, {\it outgoing modes} near the horizon are
eliminated and thus effective theory is chiral there. Then effective action
for the metric due to matter fields becomes anomalous near the horizon with
respect to general coordinate transformations. The Hawking flux of
energy-momentum tensor is determined so that it cancels the gravitational
anomaly in the {\it consistent} form at the horizon.

On the other hand, in \cite{IUW1}, \cite{IUW2} and the present paper, {\it
ingoing modes}, which are classically irrelevant to physics outside the
horizon, are integrated out near the horizon.  The Hawking fluxes are
determined by the requirement that the {\it covariant} current or
energy-momentum tensor should vanish at the horizon, instead of the
consistent current.

In the case of the Schwarzschild black hole, both of them give the same
answer.  This is because the change of the sign for the coefficients of
anomalies due to whether we consider outgoing or ingoing modes can be
cancelled by the change of sign whether we consider the consistent or
covariant form of gravitational anomalies at the horizon.  (See the
coefficients of $(f')^2$ terms in eqs. (\ref{gravanomaly1}) and
(\ref{gravanomaly2}).)

In more general cases of charged or rotating black holes, however, we have
to consider gauge currents in addition to energy flow and hence both of the
gravitational and gauge anomalies must be considered.  Since the values of
gauge anomalies at the horizon differ by a factor 2, not by their signs,
between the consistent and the covariant currents (eqs.(\ref{gaugeanomaly1})
and (\ref{gaugeanomaly2})), it cannot cancel the first change of the sign.
Because of it, when we applied the calculation \cite{RW} to charged or
rotating black holes, we could not reproduce the correct value of the
Hawking fluxes.

In the appendix in \cite{IUW2}, we have calculated the effective action for
two-dimensional free fields in charged black hole backgrounds and obtained
the form of gauge currents and energy-momentum tensor directly imposing the
regularity condition that fluxes of current or energy-momentum tensor seen
by a freely falling observer should not be singular at the horizon;
\begin{eqnarray}
  J^r &=& -\frac{e^2}{2\pi}A_t (r_H), \\
  T^r_t &=& 
   -\frac{e^2}{2\pi}A_t(r_H)A_t(r) + \frac{1}{192\pi}f'^2(r_H)
  + \frac{e^2}{4\pi}A_t^2(r_H).
\end{eqnarray}
In the cases of rotating black holes investigated in this paper, the above
current and background $U(1)$ gauge field should be regarded as those for
each $U(1)$ gauge symmetry corresponding to the diffeomorphism in the
$\phi_i$-direction.  Then it is easily found that these are equivalent to
the current $J^r_{(\phi_i),O}$ and energy-momentum tensor $T^r_{t,O}$ which
are determined imposing the condition that the covariant current should
vanish at the horizon, not the consistent current.

\end{document}